\documentclass[12pt,preprint]{aastex}

\newcommand{\kev}{keV}
\newcommand{\fe}{Fe~K$\alpha$}
\newcommand{\etal}{et al.}
\newcommand{\oeight}{\ion{O}{8}}
\newcommand{\mcg}{MCG--6-30-15}
\newcommand{\ergcms}{erg~cm$^{-2}$~s$^{-1}$}

\usepackage{graphicx}

\begin{document}

\title{X-ray Reflection from Inhomogeneous Accretion Disks: II. Emission
  Line Variability and Implications for Reverberation Mapping}

%% Use \author, \affil, and the \and command to format
%% author and affiliation information.

\author{D. R. Ballantyne\altaffilmark{1},
  N. J. Turner\altaffilmark{2} and A. J. Young\altaffilmark{3}}
\altaffiltext{1}{Canadian Institute for Theoretical Astrophysics,
McLennan Labs, 60 St. George Street, Toronto, Ontario, Canada M5S 3H8;
ballantyne@cita.utoronto.ca} 
\altaffiltext{2}{MS 169-506, Jet Propulsion Laboratory, California
Institute of Technology, Pasadena CA 91109; neal.turner@jpl.nasa.gov}
\altaffiltext{3}{MIT Center for Space Research, 77 Massachusetts
  Avenue, Cambridge MA, 02139; ayoung@space.mit.edu}

\begin{abstract}
One of the principal scientific objectives of the upcoming
\textit{Constellation-X} mission is to attempt to map the inner
regions of accretion disks around black holes in Seyfert galaxies by
reverberation mapping of the \fe\ fluorescence line. This area of the
disk is likely radiation pressure dominated and subject to various
dynamical instabilities. Here, we show that density inhomogeneities in
the disk atmosphere resulting from the photon bubble instability (PBI)
can cause rapid changes in the X-ray reflection features, even when
the illuminating flux is constant. Using a simulation of the
development of the PBI, we find that, for the disk parameters chosen,
the \fe\ and \oeight\ Ly$\alpha$ lines vary on timescales as short as
a few hundredths of an orbital time.  In response to the changes in
accretion disk structure, the \fe\ equivalent width (EW) shows
variations as large as $\sim$100~eV. The magnitude and direction
(positive or negative) of the changes depends on the ionization state
of the atmosphere. The largest changes are found when the disk is
moderately ionized. The \oeight\ EW varies by tens of eV, as well as
exhibiting plenty of rapid, low-amplitude changes. This effect
provides a natural explanation for some observed instances of short
timescale \fe\ variability which was uncorrelated with the continuum
(e.g., Mrk~841). New predictions for \fe\ reverberation mapping should
be made which include the effects of this accretion disk driven line
variability and a variable ionization state. Reflection spectra
averaged over the evolution of the instability are well fit by
constant density models in the 2--10~\kev\ region.
\end{abstract}

\keywords{accretion, accretion disks --- instabilities --- line:
  formation --- radiative transfer --- X-rays: general}

\section{Introduction}
\label{sect:intro}
Shortly after the prediction of relativistically broadened iron
K$\alpha$ fluorescence lines in the X-ray spectra of accreting black
holes \citep{fab89}, it was realized that if the lines were variable
then constraints on the emission geometry and basic black hole
parameters could be determined \citep{stel90,mp92}. These ideas
matured into the prospect of iron line 'reverberation mapping', which,
given a series of short time-resolved X-ray spectra following a
significant flare in the continuum, could accurately derive such
fundamental parameters as the black hole mass, spin, and location of
the X-ray source(s) \citep{rey99,yr00,rus00,gl04}. The promise of this
technique is such that iron line reverberation mapping of nearby
Seyfert~1s is one of the principal science goals of the upcoming
\textit{Constellation-X} observatory.

In recent years it has become obvious that reverberation mapping will
be more challenging than perhaps was previously thought. The
\textit{XMM-Newton} telescope has found only a handful of
relativistically broadened \fe\ lines, which indicates that these
components may be in general quite weak and low contrast
features. Moreover, the best example of a relativistic line (observed
from \mcg; \citealt{fab02}) exhibits less variability than the X-ray
continuum and shows no correlation with the observed broadband flux
(\citealt{ve01,vf04}, but see \citealt{pon04}).  These properties seem
to apply on long timescales in a wide sample of bright Seyfert~1
galaxies \citep*{wgy01,mev03}, but there also have been instances of
rapid iron line variation with only a slight continuum change (Mrk 841;
\citealt{pet02}), and where the line at least partially follows the
continuum \citep[e.g.,][]{nan97,iwa04}. Clearly, there is a wide range
of possible \fe\ variability characteristics.

The problem these observations cause with the standard reverberation
picture is with the underlying assumption that variations in the \fe\
line are directly related to changes in the observed continuum. The
observed behavior shows that this assumption cannot be valid at all
times, and that there are other effects which are equally important to
the production of the \fe\ line. One possible perturbation to the
previous reverberation calculations is photoionization of the
accretion disk surface by the incident X-rays (the work by
\citealt{rey99}, e.g., assumed only neutral reflection from the disk
outside of the innermost stable circular orbit). If the disk surface
is ionized to a certain level, then the \fe\ equivalent width (EW)
becomes anti-correlated with the continuum, in contrast to the
positive correlation expected if the disk remains neutral
\citep{nay00,bfr02,br02}. It is also possible that the X-ray continuum
incident on the disk is different from the observed one. For example,
light-bending effects could enhance the illumination \citep{mf04},
while an outflowing corona could diminish it \citep{bel99}. While
these effects may all be operating at some level, they are still
connected with the X-ray continuum, and therefore a relation (positive
or negative; strong or weak) between the \fe\ line flux and the
incident continuum is expected. In this paper, we outline a process by
which the \fe\ line may vary independently of the illuminating X-ray
flux.

For systems accreting at greater than 0.3\% of the Eddington rate, the
innermost annuli of the accretion disk, where the relativistic portion
of the iron K$\alpha$ line is produced, are likely radiation pressure
dominated \citep{ss73}. Numerical results indicate that turbulence
driven by the magneto-rotational instability (MRI; \citealt{bh91}) may
involve large density fluctuations in radiation-dominated disks if
magnetic pressure exceeds gas pressure and photons diffuse from
compressed regions \citep{tsks03,tur04}. Large density variations may
also be caused by the photon bubble instability (PBI;
\citealt{ar92,gam98}), which may develop into a series of shocks
propagating through the plasma \citep{beg01}. These results suggest
that radiation-dominated disks have time-varying density
inhomogeneities on scales smaller than the disk thickness. In the
first paper of this series (\citealt{btb04}, hereafter Paper I), we
showed that density variations in the outermost few Thomson depths can
lead to reflection spectra differing from those of uniform
slabs. Here, we continue our investigation of X-ray reflection from
inhomogeneous accretion disks by computing the evolution of the
reflection spectrum as the disk structure is changing due to the
PBI. As is shown below, the changing density structure in the disk can
affect the EW and line flux of the predicted emission lines, despite
the constant illuminating continuum.

The next section details the setup and conditions of the PBI model and
reflection calculations. We then present the results of the reflection
models, both as a function of time and time-averaged
(\S~\ref{sect:res}). The results are then discussed in
\S~\ref{sect:discuss} before we summarize and draw our conclusions in
\S~\ref{sect:concl}.

\section{Models: Setup and Calculations}
\label{sect:models}
An important timescale in X-ray reflection is the time for the
accretion disk to re-establish hydrostatic balance following a change
in illumination \citep{nk02,col03}.  The adjustment time is about
equal to the orbital period \citep*{fkr02}.  Turbulence inside the
disk results from the MRI and also has a characteristic overturn time
similar to the orbital period \citep{bh98}.  In radiation-dominated
disks, the PBI may grow faster than the orbital frequency
\citep{bs03}, and could lead to structures that vary over a small
fraction of an orbit.  Below we examine time variations in the X-rays
reflected from structures produced by the growth of photon bubbles.
The variations are computed in two steps, as in Paper I.  First,
density distributions resulting from the instability are found using a
two-dimensional, time-dependent numerical radiation
magnetohydrodynamical (MHD) calculation.  The density profile along a
single ray passing vertically through the disk is used as the input
for the second step.  We assume X-rays from a diffuse external source
strike the disk, and calculate the reflected spectrum by solving the
coupled equations of radiative transfer and ionization balance in one
spatial dimension.  The effects of changes in the density profile are
tracked by calculating the spectrum 100 times per orbital period.

The calculation of the growth of photon bubbles is similar to that
described in Paper I.  A narrow annulus of accretion disk in orbit
around a black hole of $M_{\mathrm{BH}} = 10^8 M_\odot$ at a distance
$R_{\mathrm{S}} = 20$ Schwarzschild radii ($= 40
GM_{\mathrm{BH}}/c^2$) is modeled in locally Cartesian coordinates in
two dimensions, with symmetry assumed along the orbit.  The
frequency-averaged equations of radiation MHD are integrated using the
ZEUS code \citep{sn92a,sn92b} with its flux-limited radiation
diffusion module \citep{ts01}, on a grid of $128\times 768$
uniformly-spaced zones.  The initial condition is a Shakura-Sunyaev
model with accretion rate 10\% of the Eddington limit for 10\%
radiative efficiency, and accretion stress parameter $\alpha=0.06$.
The total Thomson optical depth is 9200.  To this is added a uniform
magnetic field with pressure 1\% of the midplane radiation pressure,
inclined $12^\circ$ from horizontal. The initial state is disturbed
slightly by applying random density perturbations of up to 1\% in each
grid zone. The domain extends 1.2 Shakura-Sunyaev semi-thicknesses
either side of the midplane and has width one-sixth the height.  The
horizontal boundaries are periodic, and the vertical boundaries allow
gas, radiation and magnetic fields to flow out but not in.
Differential rotation is neglected, so there is no magneto-rotational
instability and no external energy source.  The radiation initially
present escapes through the top and bottom boundaries, and the gas
cools over time.  The flux of radiation energy leaving each face of
the disk initially is $F_{\mathrm{disk}}=7.7\times
10^{13}$~erg~cm$^{-2}$~s$^{-1}$.  During the first 0.8 orbits of the
calculation, photon bubbles grow to non-linear amplitudes and develop
into trains of shocks.  Photons then diffuse faster through the
low-density regions between the shocks and the disk flux is several
times greater than initially. By 2.35 orbits, three-quarters of the
initial radiation energy is lost, and the horizontally-averaged disk
thickness approximately halves. A density floor of 1\% of the initial
midplane value or $9.93\times 10^{-12}$~g~cm$^{-3}$ is imposed during
the calculation so that results are obtained with a reasonable amount
of computer time. The floor is applied in the regions between shocks,
reducing the overall density contrast. The dynamical time for this
radius is $t_{\mathrm{dyn}} \sim (R^3/GM_{\mathrm{BH}})^{1/2} \sim
\sqrt{8} (R/R_{\mathrm{S}})^{3/2} GM_{\mathrm{BH}}/c^3 \sim 1.2\times
10^5 (R/20\ R_{\mathrm{S}})^{3/2} (M_{\mathrm{BH}}/10^8\
\mathrm{M}_{\odot})$~s. The orbital period is $7.8\times 10^5$
seconds, or about nine days.  The density profile along a ray passing
vertically through domain center is stored every 0.01~orbits for use
in the X-ray reflection calculations.

Each density profile was trimmed so that the total Thomson depth
through the gas was about 10. The number of zones in the density cuts
varied from a minimum of 23 to a maximum of 71.  Reflection spectra
were then computed for each profile using the code described by
\citet{ros93} (see also \citealt*{brf01}). The illuminating continuum
was a power-law defined between 1~eV and 100~\kev\ with a photon index
$\Gamma=2$. The gas was allowed to come into thermal and ionization
balance before the reflection spectrum was computed (these timescales
are generally much shorter than the hydrostatic time; e.g.,
\citealt{nk02}). The following ions are included in the calculations:
\ion{C}{5} \ion{--}{7}, \ion{N}{6} \ion{--}{8}, \ion{O}{5}
\ion{--}{9}, \ion{Mg}{9} \ion{--}{13}, \ion{Si}{11} \ion{--}{15} and
\ion{Fe}{16} \ion{--}{27}. Four different illuminating fluxes
$F_{\mathrm{X}}$ were used for each density profile in order to
investigate differences due to ionization effects. These fluxes
corresponded to $F_{\mathrm{X}}/F_{\mathrm{disk}}=$0.5, 1, 4, and
8. Here, $F_{\mathrm{disk}}$ corresponds to the initial disk flux, and
$F_{\mathrm{X}}$ does not change with time. All 944 ionized reflection
models converged with no problems.

Of course, this procedure is not self-consistent, as the heating due
to the illuminating X-rays is not included in the gas dynamics.
Similarly, the internal radiation field in the MHD calculation does
not change the ionization state of the gas, nor is it included in the
outgoing spectrum.  Unfortunately, time-dependent 2-D and 3-D
calculations of an X-ray illuminated accretion disk with ionization
physics, frequency-dependent radiative transfer and MHD are still
years away.  However, we expect that the omissions described above do
not greatly affect our results.  The internal radiation field has
temperatures $\sim 10^5$~K, so neglecting its ionizing power does not
affect the high energy spectrum that is the focus of this paper.
Ignoring the effects of X-ray heating on the gas structure is more
problematic, as an optically thick skin may form on the disk surface in
hydrostatic models with $F_{\mathrm{X}}/F_{\mathrm{disk}} > 1$
\citep*[e.g.,][]{nkk00}.  This skin would then scatter and smear out
the reflection spectrum. However, the return to hydrostatic balance
after a pressure perturbation takes about an orbit, so the MRI and PBI
may cause permanent disequilibrium.  The turbulence driven by the MRI
can lead to density changes in the gas on the orbital period
\citep{tsks03} and can mix irradiated gas in to large optical depths.
The fastest photon bubble modes have growth time shorter than the
orbital period by the square root of the ratio of radiation to gas
pressure.  This estimate follows from Table~2 of \citet{bs03} if
radiation and magnetic pressures exceed the gas pressure, as in the
surface layers in a 3-D radiation-MHD calculation of a small patch of
disk that includes the generation and buoyancy of the magnetic fields
\citep{tur04}.  Furthermore in the numerical photon bubble calculation
described at the start of this section, the instability leads to a
radiative flux with horizontal component comparable to the vertical.
The horizontal flux varies across the face of the disk so the
atmosphere feels substantial differential stresses.  As is seen below,
the densities change over periods as short as a few percent of an
orbit, and the gas is out of hydrostatic balance from the time the
instability reaches non-linear amplitudes until the end of the
calculation.  These results suggest that like the disk itself, any
X-ray heated skin is inhomogeneous and time-variable.  The effects of
irradiation on a turbulent atmosphere may be worth investigating with
future calculations.

\section{Reflection Results}
\label{sect:res}
\subsection{Time-dependent Variations} 
\label{sub:orbit}
\begin{figure}
\centerline{
\plottwo{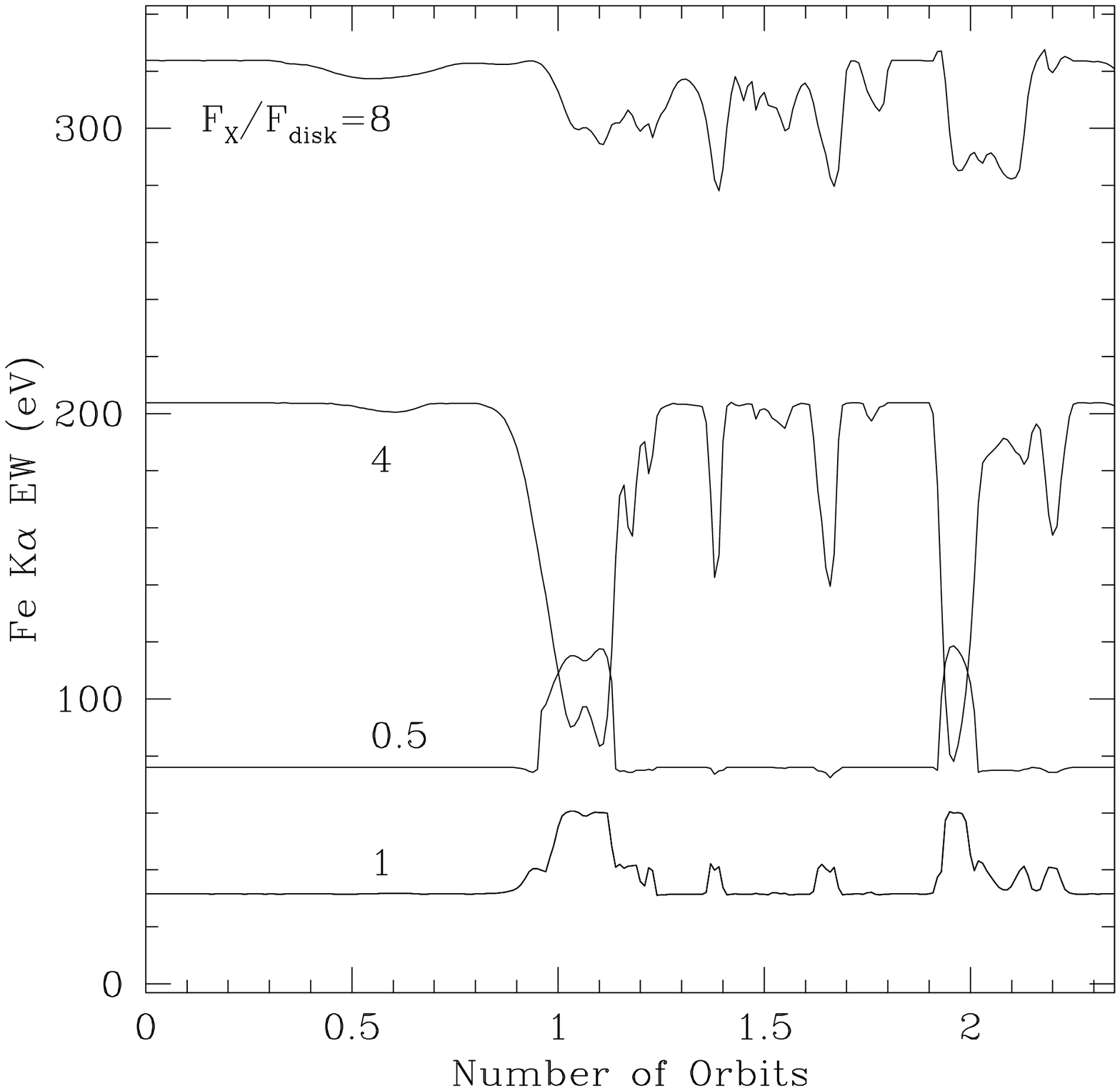}{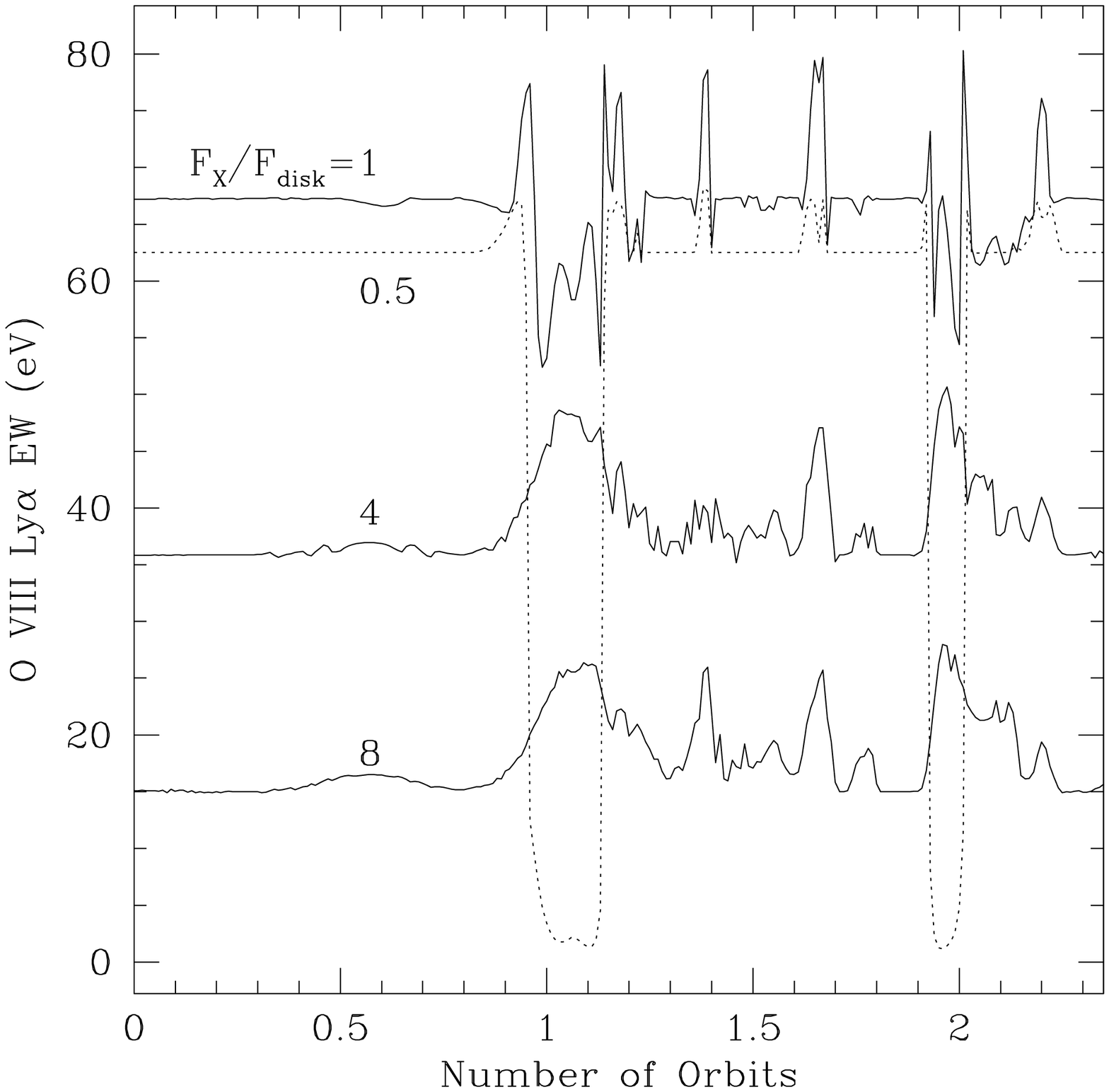}
}
\caption{The evolution of the \fe\ and \oeight\ Ly$\alpha$ equivalent
  width (EW) as the photon bubble simulation runs over 2.35
  orbits. The $\Gamma=2$ power-law was added to the reflection spectra
  before the EW was calculated. The \fe\ EW exhibits variability over
  timescales as small as a few one-hundredths of an orbit, and can
  change by factors up to 2.5. The large drop of the \oeight\ EW when
  $F_{\mathrm{X}}/F_{\mathrm{disk}}=0.5$ is due to increases in
  density causing significant absorption of the reflection spectrum.}
\label{fig:ews}
\end{figure}

To quantify any changes made to the reflection spectra by the
evolution of the PBI, we calculated the \fe\ and \oeight\ Ly$\alpha$
EWs from each spectrum. Since the $\Gamma=2$ power-law will also be
observed along with the reflection component, the two spectra were
summed before the EW calculations were performed. The results are
shown in Figure~\ref{fig:ews}, with the different lines denoting the
different values of $F_{\mathrm{X}}/F_{\mathrm{disk}}$. The \fe\ EW
spans the range from 20--60~eV when
$F_{\mathrm{X}}/F_{\mathrm{disk}}=$1 and the line is suppressed due to
Auger destruction \citep*{rfb96}, up to $\sim 320$~eV when
$F_{\mathrm{X}}/F_{\mathrm{disk}}=$8 and a strong ionized line at
6.7~\kev\ is present. The plots show that there was little variation
in the EWs of the emission lines over the first orbit, but large
changes occurred afterward as the PBI grew in magnitude. A very
significant variation in the \fe\ EW occurred between orbits 0.9 and
1.2, where the density at the outermost layers of the atmosphere
increased by $\sim 50$\%. Despite the small increase in density, it
greatly altered the \fe\ EW, particularly when the line originated
from ionized iron (i.e., when
$F_{\mathrm{X}}/F_{\mathrm{disk}}=$4). In that case, the \fe\ EW
dropped by a factor of $\sim2.5$ due to a decrease of the effective
ionization parameter and an increase of the continuum absorption. On
the other hand, when a neutral 6.4~\kev\ line dominated or when the
atmosphere was very highly illuminated, the small increase in density
between orbits 1.0 and 1.1 caused much smaller changes in the \fe\ EW
(20--40~eV).

As mentioned in \S~\ref{sect:models}, a density floor equivalent to
$4.25\times 10^{12}$~cm$^{-3}$ was used in the PBI simulation. Tests
with a floor artificially lowered by a factor of 10 resulted in a more
ionized reflection spectrum, as if $F_{\mathrm{X}}$ had been increased
and the density was left unchanged. We conclude that the density floor
will not qualitatively affect the results presented in Fig.~\ref{fig:ews}.

\begin{figure}
\centerline{
\plottwo{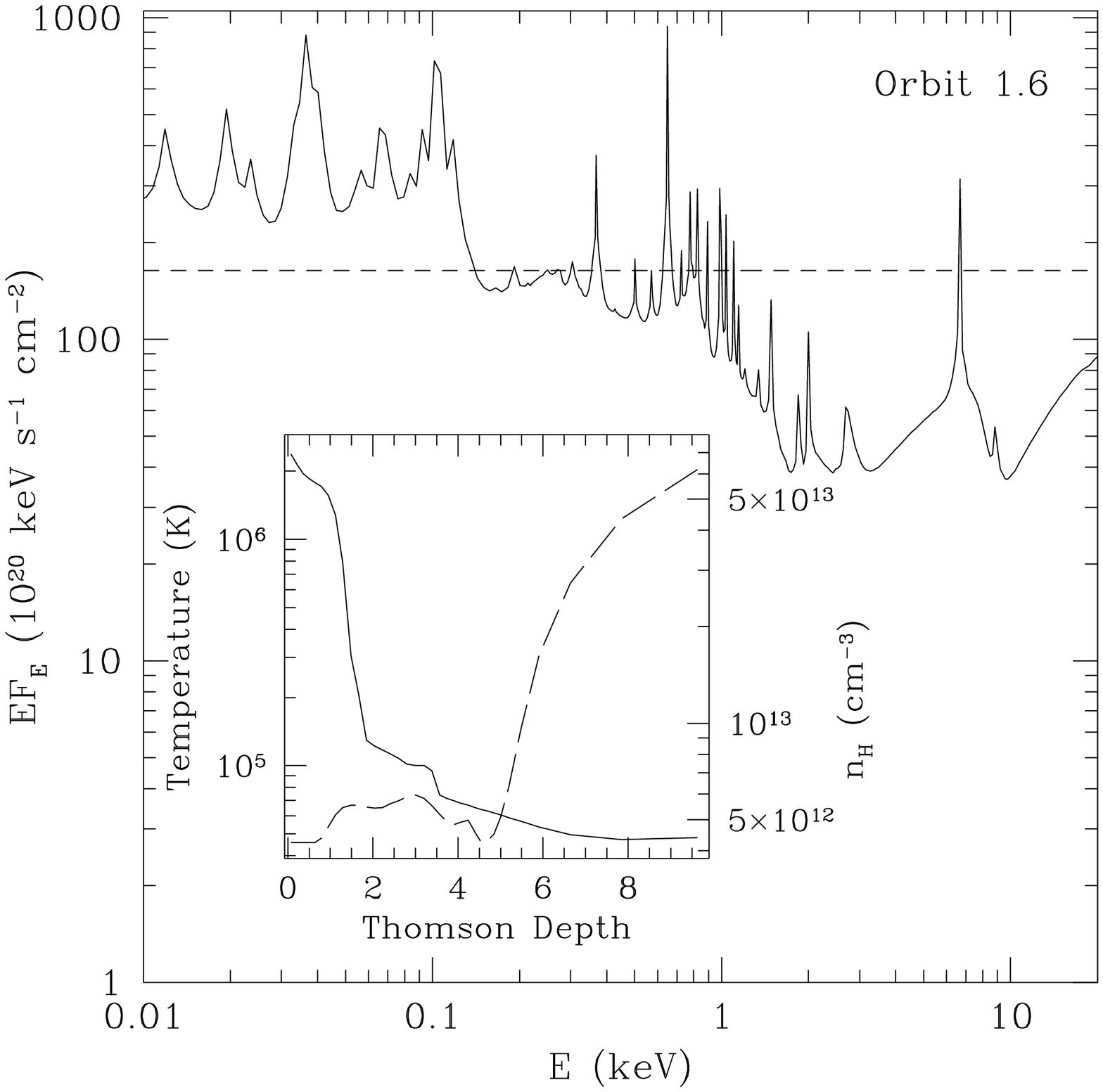}{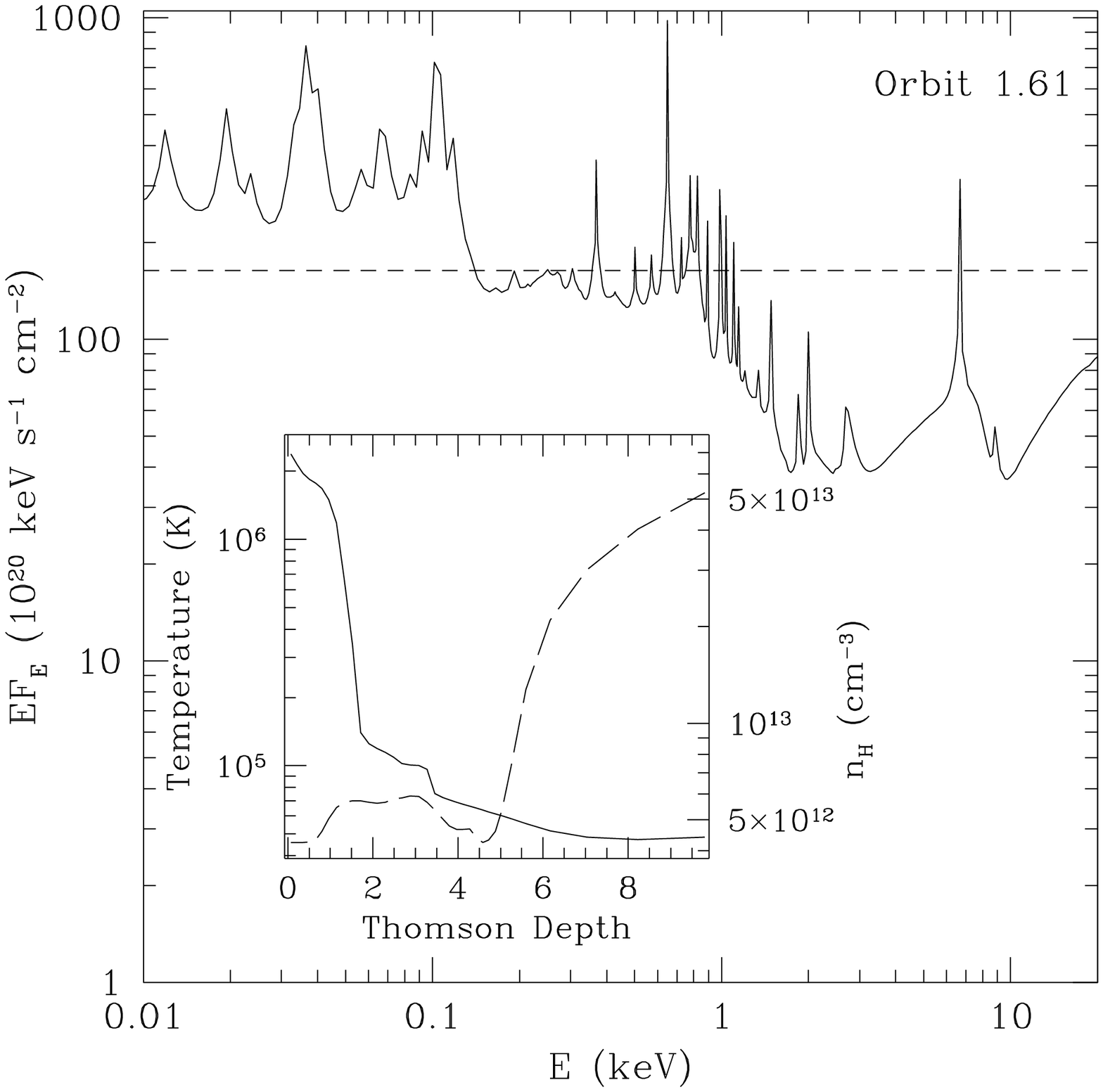}
}
\centerline{
\plottwo{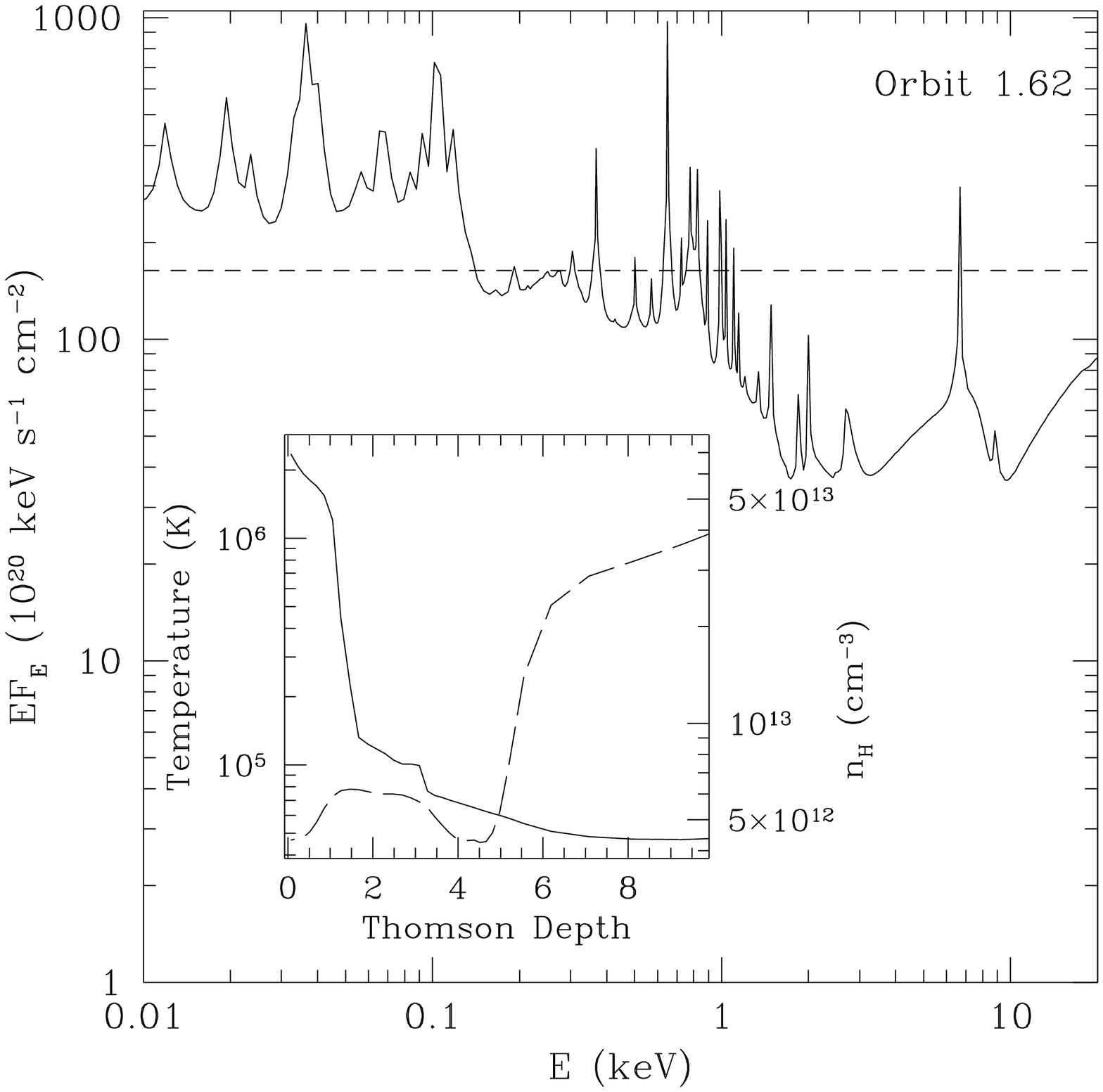}{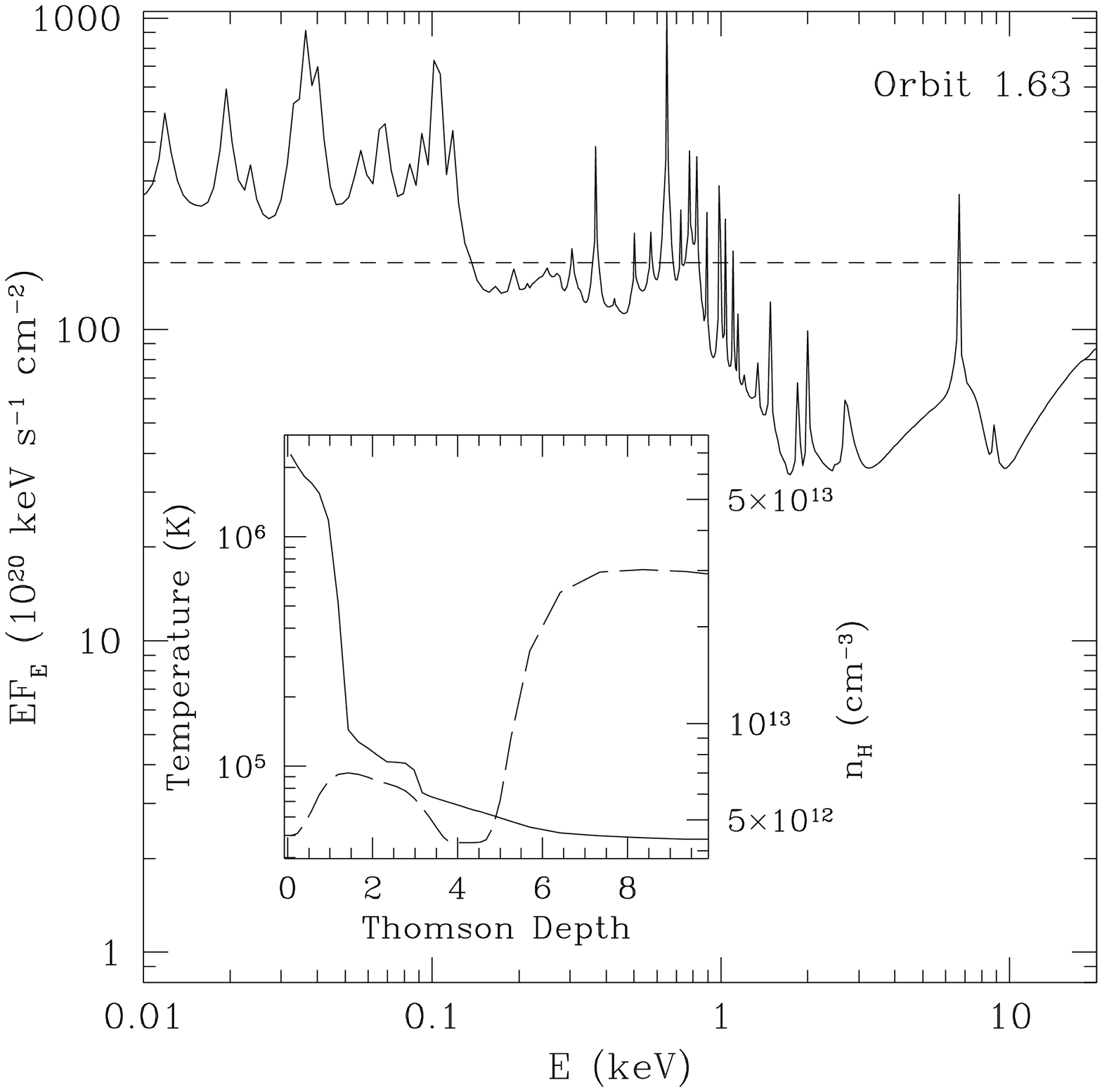}
}
\caption{X-ray reflection spectra (solid line) calculated from the
outer 10 Thomson depths of a photon bubble simulation when
$F_{\mathrm{X}}/F_{\mathrm{disk}}=4$. The short-dashed
line denotes the $\Gamma=2$ power-law that was incident on the
material. The insert shows the gas temperature (solid line) and number
density (dashed line) for each model. The different plots show how the
reflection spectra, temperature and density changed as the simulation
evolved from Orbit 1.6 to Orbit 1.69.}
\label{fig:orbits}
\end{figure}

As an illustration of the rapidity of the possible changes to the
reflection spectrum, Figure~\ref{fig:orbits} plots the model spectra
between 0.01~\kev\ and 20~\kev\ for every 0.01 of an orbit between
orbits 1.6 and 1.69. The illuminating flux is
$F_{\mathrm{X}}/F_{\mathrm{disk}}=$4, and, as seen in
Fig.~\ref{fig:ews}, the \fe\ EW drops and recovers by $~\sim60$~eV
over this range. The inset in each panel shows the hydrogen number
density $n_{\mathrm{H}}$ profile used in the reflection calculation
and the equilibrium gas temperature found from the model. These plots
illustrate that the density profile in the outer 10 Thomson depths of
the accretion disk can change significantly over as little as 0.01 of
an orbit. Some of the changes affect the reflection spectra, and some
of them do not (depending on how strongly the atmosphere is
photoionized). However, in this case, a rapid increase in density at
the surface of the atmosphere caused a significant variation in the
strength of the predicted \fe\ line.

The second region of large amplitude line variability takes place
about a full orbit after the first instance, between orbits 1.9 and
2.2. The density at the outer part of the atmosphere increased rapidly
(over only 0.06 orbits) as a shock front propagated diagonally along
magnetic field lines into the ray we are considering. The resulting
changes in density structure as the shock passes through the ray
results in the \fe\ EW to vary drastically and show slight oscillatory
behavior ($F_{\mathrm{X}}/F_{\mathrm{disk}}=$1). We have thus clearly
shown that changes in the accretion disk structure can cause rapid
variability in the \fe\ EW, \emph{despite the continuum remaining
constant}.

The right-hand panel in Figure~\ref{fig:ews} illustrates the changes
to the \oeight\ Ly$\alpha$ line resulting from the evolution of the
PBI. The line, of course, is weaker when the disk is more highly
ionized at large $F_{\mathrm{X}}/F_{\mathrm{disk}}$. The density
enhancements that pass through the surface of the disk after about 1
orbit, decrease the local ionization parameter and increase the
\oeight\ EW, often by $>10$~eV, except for
$F_{\mathrm{X}}/F_{\mathrm{disk}}=$1 and 0.5. In these cases the
increase in density makes it difficult for the weak continuum to
ionize much oxygen, resulting in enhanced continuum absorption that
suppresses that low energy reflection spectrum. In these cases, the
\oeight\ EW will be (sometimes drastically) reduced. Unlike the \fe\
line, the \oeight\ EW exhibits small amplitude rapid
variability between orbits 1 and 2. This is because the soft X-ray
lines are very sensitive to where the illuminated gas is most rapidly
cooling. As a result, the strength of the \oeight\ line, as well as
the \ion{N}{7} and \ion{C}{6} lines, are more sensitive to the
density structure than the \fe\ line (see also Paper I).

Since the EW is a measure of the line strength relative to the
continuum, and the illuminating continuum is held constant in these
calculations, the variations in both the \fe\ and \oeight\ EW are
entirely due to intrinsic changes in the line flux in the
reflection spectrum. Therefore, measurements of the line flux would
also see changes of similar magnitude due to this effect.

\subsection{Time Averaged Spectra}
\label{sub:avg}
The variations presented above occurred on timescales as short as a
few one-hundredths of an orbit, or $\sim 5$--$10$~ks for the radius
and black hole mass assumed for the simulation. This time is too short
to accumulate a high quality spectrum with modern observatories such
as \textit{XMM-Newton} and \textit{Chandra} except for the brightest
Seyfert~1 galaxies. Typical observations need to be many times longer
for proper spectral analysis, which will average over any variations
in the emission lines due to the changing accretion disk
structure. Thus, it is interesting to check whether the time-averaged
spectra presented in \S~\ref{sub:orbit} retain any knowledge of the
unstable disk structure.

\begin{figure}
\centerline{
\includegraphics[angle=-90,width=0.9\textwidth]{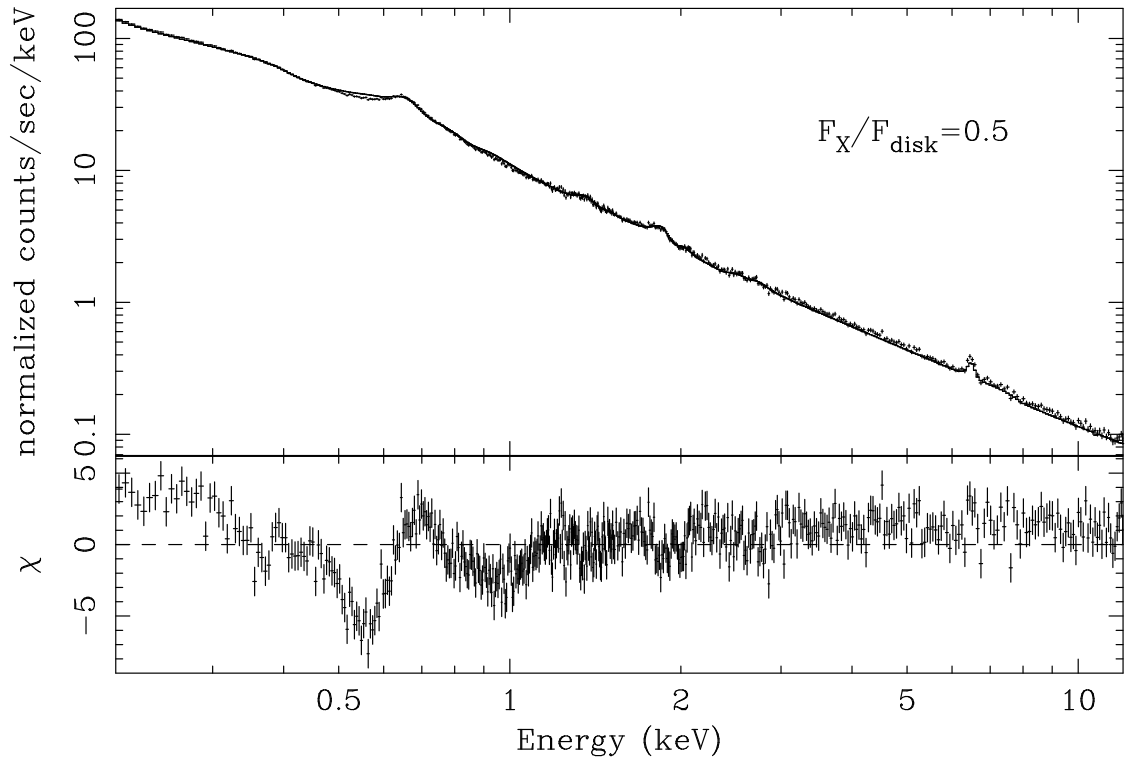}
}
\caption{Simulated count spectra and residuals (in units of standard
  deviations) of the 0.2--12~\kev\ constant density fit to the time averaged
  $F_{\mathrm{X}}/F_{\mathrm{disk}}=$0.5 model. The fit parameters are
  given in Table~\ref{table:avg}.}
\label{fig:resid}
\end{figure}

We follow very similar procedures as described in Paper I. For each
value of $F_{\mathrm{X}}/F_{\mathrm{disk}}$, we averaged together 156
reflection spectra between orbits 0.8 and 2.35 (in order to
concentrate on the region where the PBI is most important) and
simulated a 40~ks \textit{XMM-Newton} observation\footnote{Strictly
speaking, a 780~ks observation would be needed to observe a full
orbit of this PBI simulation. However, since the normalization of the
model is chosen so that Poisson noise never dominates, the actual
length of the simulated observation does not change the results.} with
XSPEC v.11.3.0p \citep{arn96}.  The $\Gamma=2$ power-law was added to
each averaged reflection spectrum so that the reflection fraction $R$
is unity. The simulated spectra were then fit with reflection models
computed by the same code assuming a constant density of
$n_{\mathrm{H}}=10^{13}$~cm$^{-3}$.  The results of the spectral
fitting are shown in Table~\ref{table:avg}.  The fits are generally
acceptable in the 0.2--12~\kev\ energy range, but are greatly
improved in the 2--10~\kev\ band. As a result of their sensitivity
to the density structure, the soft X-ray emission features caused most
of the difficulty in fitting the 0.2--12~\kev\ simulated data (see
also Paper I). The worst reduced $\chi^2$ obtained was for the
$F_{\mathrm{X}}/F_{\mathrm{disk}}$=0.5 model over the 0.2--12~\kev\
band. The residuals to this fit are shown in Figure~\ref{fig:resid},
and are due to the slab model incorrectly accounting for the soft
X-ray emission lines. However, the high energy continuum is well
modeled by a uniform reflector. This exercise suggests that the
density variations at the surface of the disk will not greatly affect
the time-averaged reflection spectra above 2~\kev\ (but see
\S~\ref{sub:impli}). Moreover, constant density reflection models
will be useful to parameterize observed spectra with an
ionization parameter and reflection strength provided that the
observing time is long compared to the fluctuation timescale in the
disk.

\section{Discussion}
\label{sect:discuss}

\subsection{Variable Reflection Features}
\label{sub:var}
In Paper I we showed that density inhomogeneities in the surface
layers of radiation dominated accretion disks could potentially impact
the reflection spectrum. That paper also suggested that the motion and
evolution of any inhomogeneities in a real accretion disk might cause
line variability that is independent of the illuminating radiation
field. Here, we have confirmed that suggestion by explicitly showing
variations in the \fe\ and \oeight\ Ly$\alpha$ EW on timescales as
short as a few one-hundredths of an orbital time as a result of the
evolution of the PBI, which is relevant for radiation dominated
accretion disks. This mechanism provides a natural explanation for any
short timescale \fe\ variability that is uncorrelated with the
continuum \citep[e.g.,][]{nan99,wwz01}. In particular,
Fig.~\ref{fig:ews} indicates that the sudden change in \fe\ EW
observed in Mrk~841 \citep{pet02} could easily be accounted for by
changes in the accretion disk structure. Furthermore, as mentioned in
Paper I, the narrow, redshifted lines that have been recently inferred
in the spectra of NGC~3516 \citep{ttj02} and Mrk~766 \citep{tkr04}
could originate in density inhomogeneities at a particular radius in
the disk atmosphere. If this is the case, then our results indicate
the lines should be variable on timescales about 1\% of the orbital
period of the emitting patch. While the statistics are generally poor,
there is evidence for variability in these narrow lines. Future data
will be able to provide a better test of this hypothesis.

The time-averaged reflection spectra from the PBI calculation showed
no obvious indication of the variations on smaller
timescales. However, we cannot rule out rapid variations in disk
structure that produce iron K$\alpha$ changes drastic enough to be
detected in observations lasting several orbits. In this paper, we
considered only one black hole mass, accretion rate, position in the
disk and magnetic field arrangement.  For the parameters chosen, the
\fe\ region is readily fit with a uniform-slab reflection model, while
the soft X-ray region in some cases is not. Clearly, there is wide
parameter space that needs to be explored in order to determine the
full impact the PBI and MRI will have on X-ray reflection spectra.

The results in \S~\ref{sect:res} have important implications for
upcoming \fe\ reverberation mapping attempts with
\textit{Constellation-X}. If accretion disk structural changes can
cause the \fe\ line to vary independently of the continuum on
timescales much smaller than $t_{\mathrm{dyn}} \sim \sqrt{8}
(R/R_{\mathrm{S}})^{3/2} GM_{\mathrm{BH}}/c^3$ ($\sim 1.2\times
10^5$~s, for the photon bubble simulation shown here), then it will
become an important source of noise for the reverberation
experiments. This will be especially true for reverberation
experiments probing small radii around low mass black holes (say,
$M_{\mathrm{BH}} \sim 10^{6-7}$~M$_{\odot}$). In that case, the disk
structure may vary on timescales as short as 100 seconds or less,
comparable to the X-ray variability timescale
\citep[e.g.,][]{vfn03}. Furthermore, the current transfer functions
computed for reverberation mapping have all neglected the effects of
ionization of the disk surface \citep[e.g.][]{rey99}, which can result
in an anti-correlation between the line EW and continuum
\citep{br02}. Therefore, it is possible that the combined effects of a
variable density structure (due to the PBI and perhaps other
instabilities) and ionization state (due to changes in the X-ray
illumination) will significantly complicate the results of any
reverberation experiment. What is needed to provide concrete
predictions for reverberation mapping is time dependent ionized
reflection calculations where the illuminating flux is related to the
underlying magnetic stress \citep[e.g.,][]{ar03} or outgoing magnetic
flux \citep{ms00}. This is a direction for future research.

In the standard reverberation mapping scenario, the idea is to observe
changes in the \fe\ line, as X-rays from a flare sweep across the
disk. The PBI will likely be ongoing at every radius in the inner
region of the disk, but will be subject to local conditions, such as
the strength and geometry of the magnetic field. Moreover, the linear
growth rate of the PBI increases as one moves to smaller radii
\citep{bs01}. Therefore, the disk structure will be very position
dependent as the X-rays flash overhead, so that the reflection
spectrum could show variability due to disk inhomogeneities as a
function of position, as well as time. PBI simulations spanning a
larger range in radius would be necessary to test the possibility of
reflection spectra variability due to positional changes.

If different radii produce different variability properties, than the
observed variations may be diluted if the X-rays illuminate a wide
enough range of radii. However, there are often times when the AGN
continuum is likely dominated by a single flare or a number of small
flares which occur in close proximity to one another
\citep[cf.][]{mf01}. In this case, which is the most relevant for
reverberation mapping, then the one-zone models employed here may be
useful.

The signature of a changeable disk atmosphere is, of course, short
timescale \fe\ (and \oeight\ Ly$\alpha$, if observable) variability
which is independent of the observed continuum. Therefore, it may be
possible to observationally test if this process is important. As
mentioned above, such cases of rapid changes in the \fe\ line have
been observed (e.g., Mrk~841), but these cannot be used as evidence
that effects of disk instabilities have been observed (although it may
provide a good explanation). There still remains too many unknowns
regarding the location and dynamics of the X-ray source providing the
illuminating continuum. The long timescale disconnectedness between
the \fe\ line and continuum \citep[e.g.,][]{mev03} suggests it is
possible, if not likely, that at some level the disk sees a different
irradiating continuum than what is observed. Numerical simulations of
coronal formation above accretion disks, still some years away, may
provide the clues to understanding the X-ray source in AGN.

\subsection{Implications of the Time Averaged Results}
\label{sub:impli}
The reflection spectra that were averaged over the variability caused
by the PBI were well described by a constant density model in the
2--10~\kev\ band. The implication is that, over this energy range, the
spectrum is dominated by emission from a single ionization parameter.
Since the illuminating flux is known for these test cases, the best
fit $\xi$ can be used to determine $\tilde{n}_{\mathrm{H}}=4 \pi
F_{\mathrm{X}}/\xi$, the density at which the most important spectral
features are formed (see Table~\ref{table:avg}). The value of
$\tilde{n}_{\mathrm{H}}$ is always less than 10$^{13}$~cm$^{-3}$, the
density actually used in the constant density models, indicating that
the ionization level is higher than 'expected' given the illuminating
flux. As can be seen in the density profiles plotted in
Fig.~\ref{fig:orbits}, $\tilde{n}_{\mathrm{H}}$ predominantly falls
within the first 1--2~Thomson depths of the atmosphere. We conclude that
reflection spectra above 2~\kev\ are effectively determined by the
ionization parameter at a Thomson depth of $\sim 1$. At lower
energies, the carbon, nitrogen and oxygen cooling lines, which are
very sensitive to the density structure, cause the spectra to exhibit
features from different effective ionization parameters, and thus are
more difficult to fit with a uniform density model.  

The fits with the constant density models suggest that changes in the
calculated spectra on 0.01-0.1 orbit timescales are difficult to
detect in the longer integrations.  However, there may be regimes where
rapid changes in disk structure produce clear signatures in the Fe
K$\alpha$ region in observations lasting several orbits.  As mentioned
in \S~\ref{sub:var}, we considered only one black hole mass, accretion
rate, position in the disk and magnetic field arrangement. Thus, there
is a wide parameter space to be explored to determine the impact of
accretion disk inhomogeneities on X-ray reflection
spectra. Nevertheless, insofar as can be determined, these results
(along with those from Paper I) indicate that constant density models
provide a good indication of the average ionization state of the accretion
disk atmosphere on long timescales.

\section{Conclusions}
\label{sect:concl}
Our conclusions from the second paper in our study on X-ray reflection from
heterogeneous accretion disks can be summarized as follows:

\begin{enumerate}
\item Changes in the density structure of an accretion disk atmosphere
  from the photon bubble instability can give rise to rapid
  variability of emission lines in the X-ray reflection
  spectrum. These variations would be observed to be independent of
  the illuminating continuum.

\item This process gives a natural explanation for any short timescale
  (e.g., hours) line variability that seemed to be uncorrelated with
  the X-ray continuum (for example, Mrk~841). However, these
  observations cannot be used as proof that disk instabilities are
  causing the line variations.  Nor can it offer an explanation for the
  uncorrelated behavior of the \fe\ line on longer timescales (greater
  than a day), although we can not rule out such an effect in a
  different region of parameter space. 

\item The timescale of the variability was found to be as small as a
  few one-hundredths of an orbital time. Therefore, this effect,
  combined with any changes in ionization state, could be a serious
  source of noise for future reverberation studies.

\item Averaging the individual reflection spectra over the 1.55 orbits
  most affected by the PBI yielded spectra corresponding to the more
  common long timescale X-ray AGN observations. These averaged spectra
  were well fit above 2~\kev\ by constant density models.

\item Photon bubble structures move across the surface of the disk in the
   radiation MHD calculations, passing in and out of the ray where we
   extracted density profiles.  Variations in the X-rays reflected
   from a region containing many structures may be less than those
   from a single line of sight.  The separations, strengths and
   orientations of the structures are found to vary with the alignment
   and strength of the magnetic fields.  The fields are likely to
   evolve over orbital timescales along with the turbulence inside the
   disk.  Two and three-dimensional reflection calculations may be
   useful in studying the consequences for reverberation mapping.
\end{enumerate}

\acknowledgments

The authors thank Omer Blaes for useful discussions, and the referee
for helpful comments which improved the paper.  DRB acknowledges
financial support by the Natural Sciences and Engineering Research
Council of Canada. NJT was supported by a National Research Council
fellowship at the Jet Propulsion Laboratory, California Institute of
Technology. AJY was supported by NASA through contract
NAS8-01129. This work was partly supported by U. S. National Science
Foundation grant AST-0307657.

\clearpage

\begin{deluxetable}{cccccccccc}
\tabletypesize{\small}
\rotate
\tablewidth{0pt}
\tablecaption{\label{table:avg}Results of fitting the
  time averaged photon bubble reflection spectra with constant density models.}
\tablehead{
\colhead{} & \multicolumn{4}{c}{0.2--12~\kev} &
\multicolumn{4}{c}{2.0--10~\kev}\\
\colhead{$F_{\mathrm{X}}/F_{\mathrm{disk}}$} &
\colhead{$\Gamma$} & \colhead{$\xi$} & \colhead{$R$} & \colhead{$\chi^2$/d.o.f.} &
\colhead{$\Gamma$} & \colhead{$\xi$} & \colhead{$R$} &
\colhead{$\chi^2$/d.o.f.} & {$\tilde{n}_{\mathrm{H}}$}}
\startdata
0.5 & 2.012$\pm 0.002$ &
126$\pm 0.3$ & 1.021$^{+0.014}_{-0.011}$ & 3832/2358
& 2.017$^{+0.013}_{-0.012}$ & 74$^{+13}_{-10}$ &
0.891$^{+0.101}_{-0.091}$ & 1717/1595 & 6.5$\times 10^{12}$\\
1 & 2.005$^{+0.003}_{-0.002}$ &
249$\pm 10$ & 0.993$^{+0.013}_{-0.012}$ & 2706/2358
& 2.008$\pm 0.010$ & 219$^{+51}_{-36}$ & 0.889$^{+0.112}_{-0.109}$ &
1714/1595 & 4.4$\times 10^{12}$\\ 
4 & 1.977$^{+0.003}_{-0.002}$ & 697$^{+10}_{-11}$ &
1.109$^{+0.020}_{-0.019}$ & 3814/2358 & 2.022$^{+0.010}_{-0.011}$ &
964$^{+50}_{-43}$ & 0.856$^{+0.061}_{-0.063}$ & 1615/1595 & 4$\times
10^{12}$\\ 
8 & 1.997$^{+0.002}_{-0.001}$ & 1122$^{+0.3}_{-18}$ &
0.837$^{+0.010}_{-0.017}$ & 3432/2358 & 2.024$^{+0.008}_{-0.010}$ &
1738$^{+185}_{-182}$ & 0.912$^{+0.047}_{-0.049}$ & 1554/1595 &
4.45$\times 10^{12}$\\ 
\enddata
\tablecomments{Averaged between 0.8 and 2.35 orbits. The normalization
  of the photon bubble models were chosen so that the 'observed'
  2--10~\kev\ flux was 3--4$\times 10^{-8}$~\ergcms. $R$ is the
  reflection fraction defined as total=incident+$R\times$reflected,
  and d.o.f.=degrees of freedom.  $\xi=4\pi
  F_{\mathrm{X}}/n_{\mathrm{H}}$ is the ionization parameter and has
  units of erg~cm~s$^{-1}$. $\tilde{n}_{\mathrm{H}}$ is the density
  derived from the known $F_{\mathrm{X}}$ and fitted $\xi$. It has
  units of cm$^{-3}$. The error-bars are the 2$\sigma$
  uncertainty for the parameter of interest. The simulated data were
  constructed using the \textit{XMM-Newton} response matrix
  \texttt{epn\_sw20\_sdY9.rmf} and assumed an exposure time of 40~ks.}
\end{deluxetable}

\clearpage

%% Use the figure environment and \plotone or \plottwo to include 
%% figures and captions in your electronic submission.

\clearpage

\clearpage
                                                                               
\begin{figure}
\figurenum{\ref{fig:orbits} \emph{cont}}
\centerline{
\plottwo{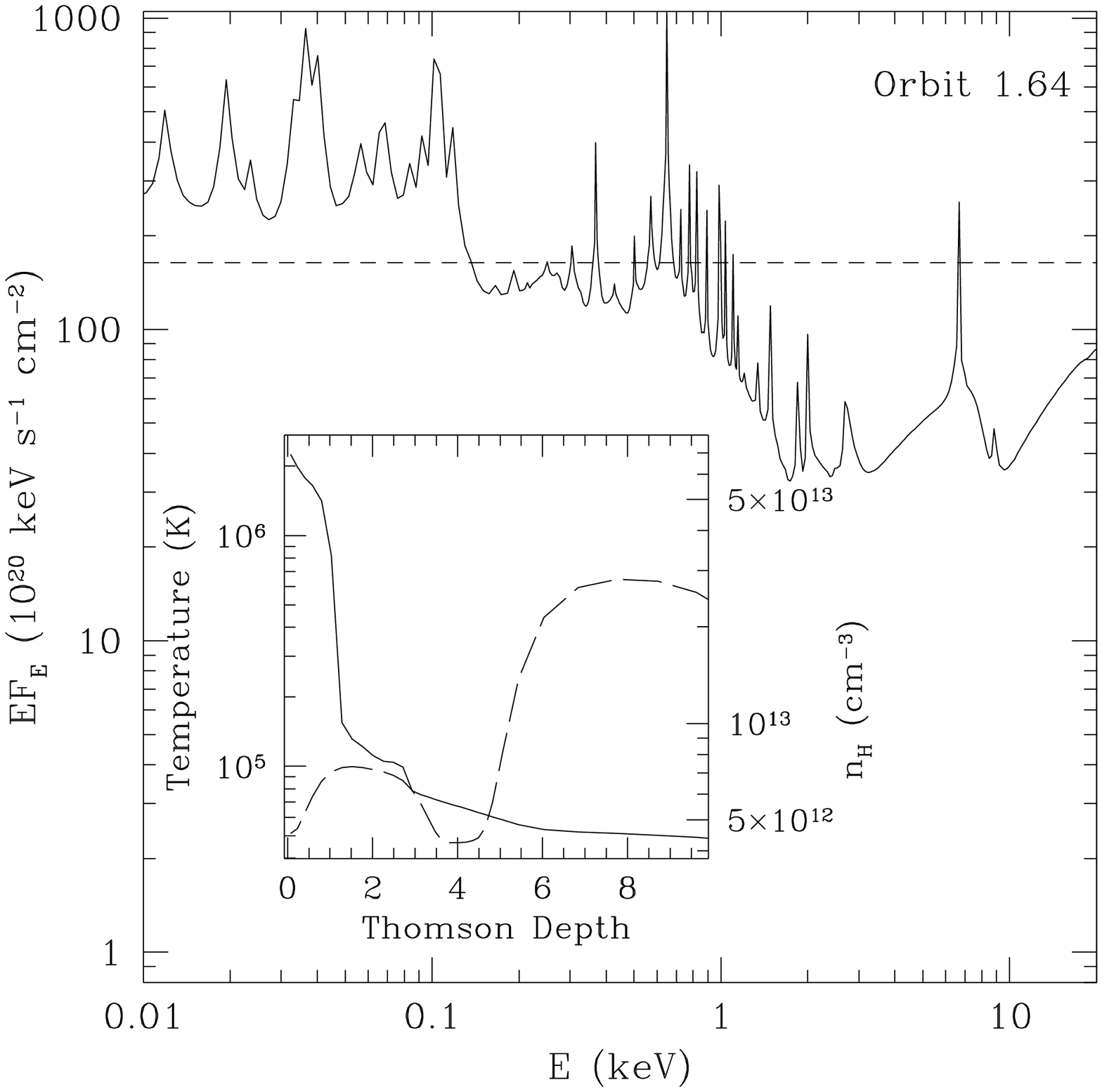}{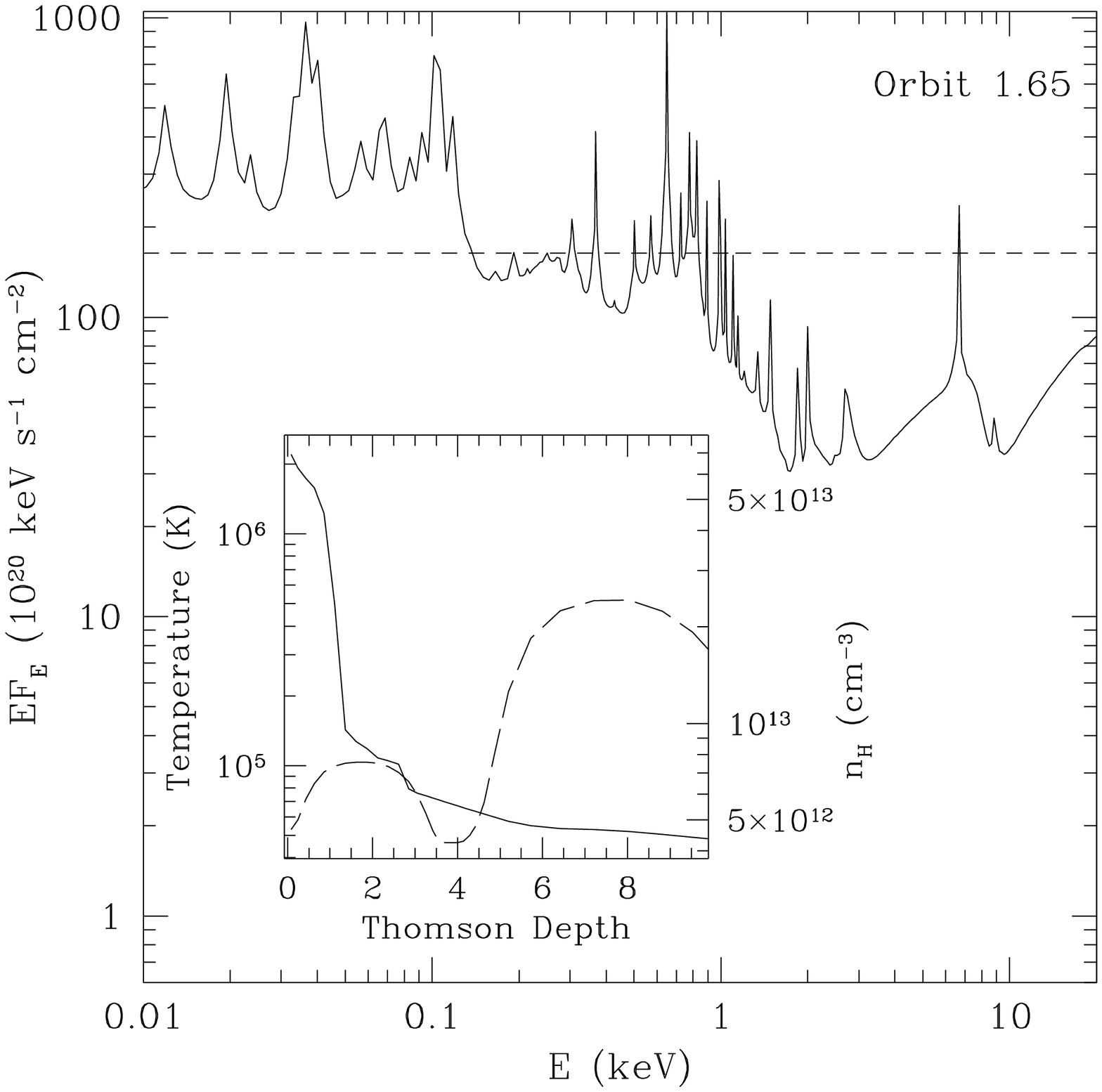}
}
\vspace{-0.5cm}
\centerline{
\plottwo{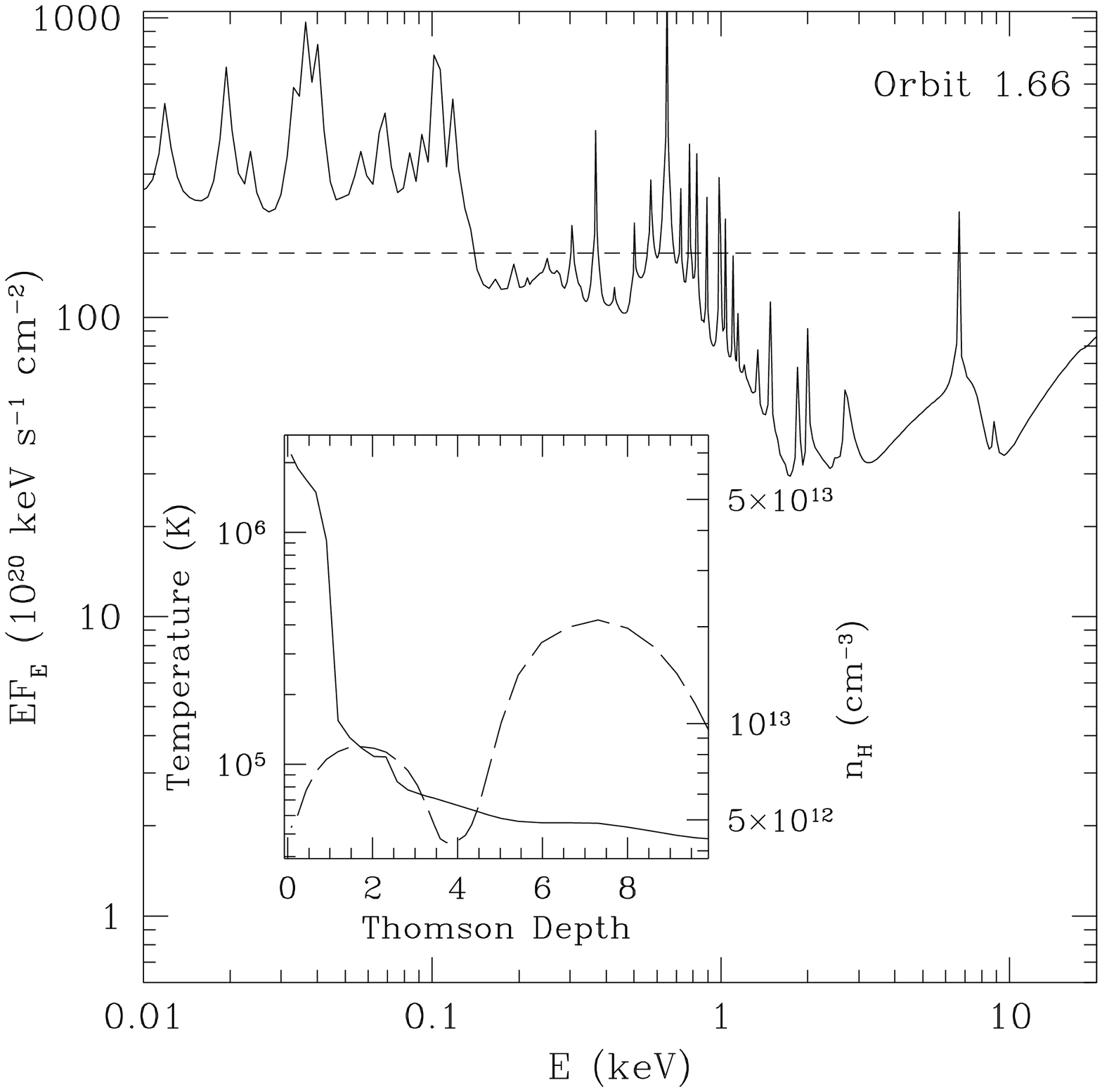}{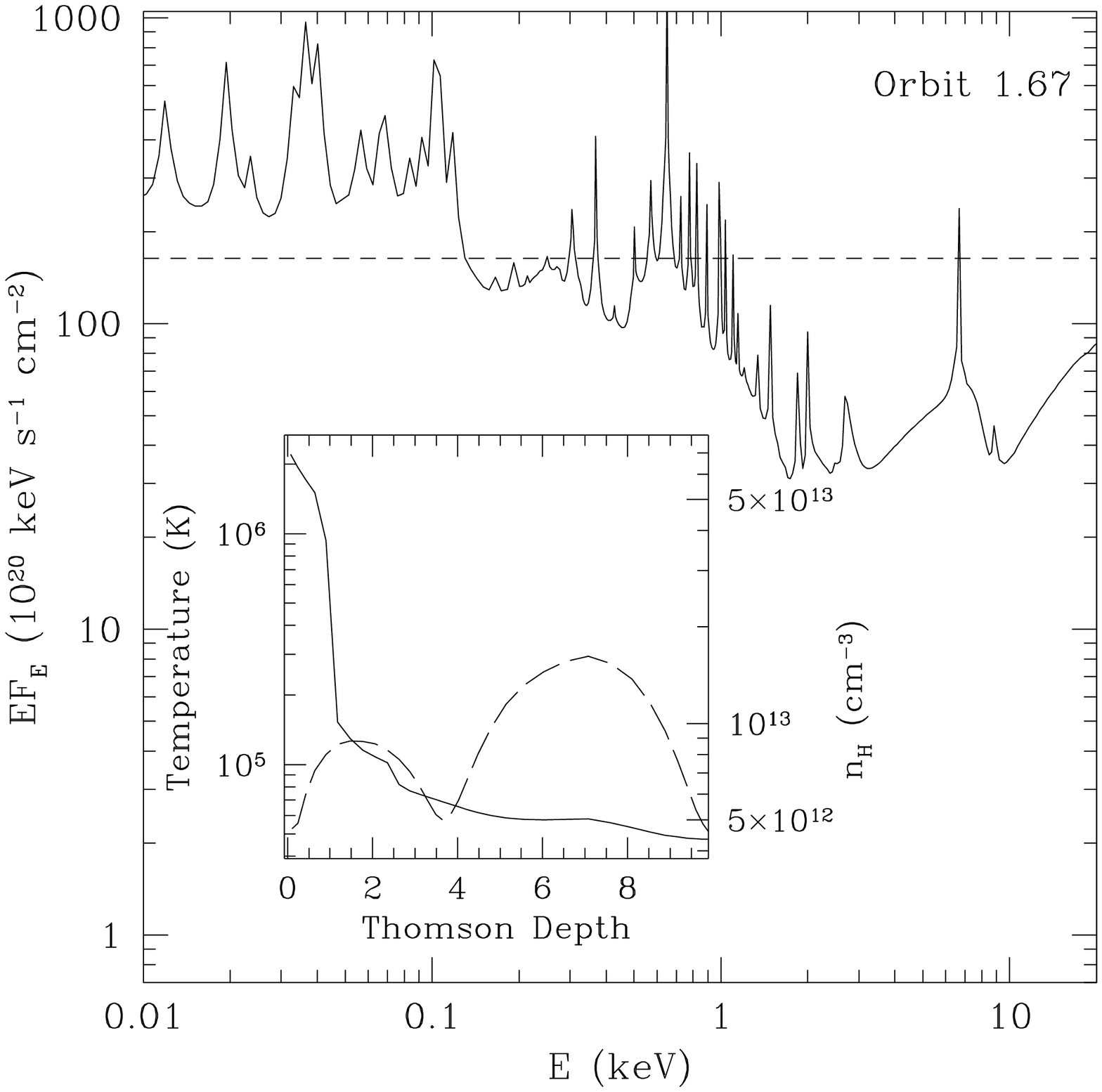}
}
\vspace{-0.5cm}
\centerline{
\plottwo{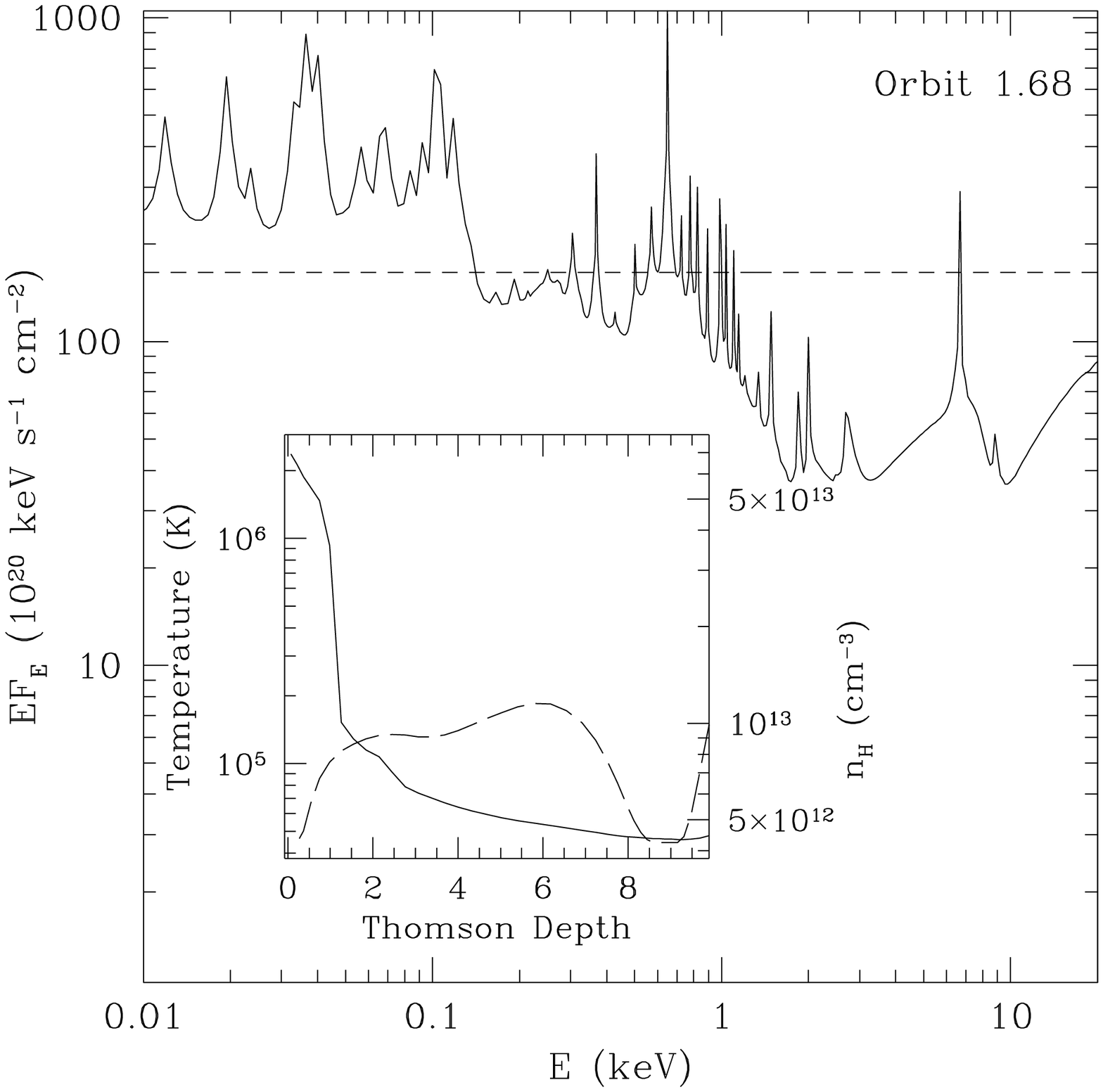}{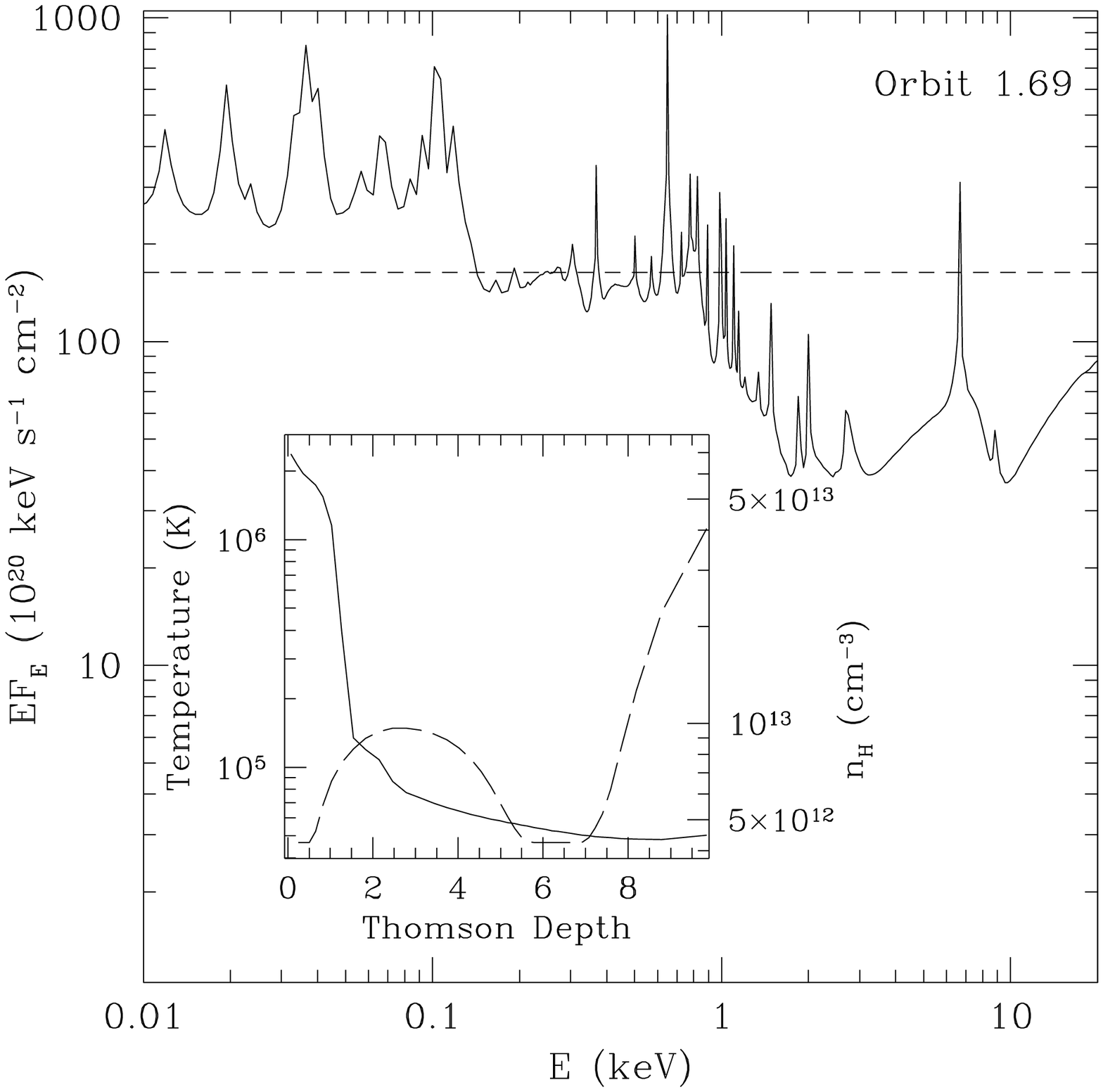}
}
\caption{}
\end{figure}

\clearpage

\clearpage

\end{document}